\documentclass{article}
\def\eq{\!\!\!\! &=& \!\!\!\! }
\def\br{\begin{eqnarray}}
\def\er{\end{eqnarray}}
\def\be{\begin{equation}}
\def\ee{\end{equation}}

\def\>{\rangle}              
\def\<{\langle}              
\def\({\left(}
\def\){\right)}
\def\[{\left[}
\def\]{\right]}

\def\Ap{{\cal A}^{+}}
\def\Am{{\cal A}^{-}}
\def\nn{\nonumber}
\def\ni{\noindent}
\def\br{\begin{eqnarray}}
\def\er{\end{eqnarray}}
\def\be{\begin{equation}}
\def\ee{\end{equation}}
\newcommand{\sss}{\vspace{.2in}}
\newcommand{\sms}{\vspace{.1in}}

\begin{document}

\centerline{\LARGE\bf Supersymmetry in the Half-Oscillator - Revisited}

\vspace*{.25in}
\begin{center}{\Large Asim Gangopadhyaya\footnote{agangop@luc.edu} and Jeffry V. Mallow\footnote{jmallow@luc.edu}} \\
Loyola University Chicago, Chicago, IL 60626
\end{center}

\vspace*{.5in} \centerline{\large{\bf Abstract}}Following a recent study by Das and
Pernice\cite{das}, we have carefully analyzed the  half-harmonic oscillator. In
contrast to their observations, our analysis reveals that the spectrum does not allow
for a zero energy ground state and hence the supersymmetry is broken when the domain is
restricted to the positive half of the real axis.

\sss \noindent {\large{\bf Introduction:}}

The harmonic oscillator is an extensively studied problem in classical and quantum
physics. This was the first problem to be studied by the ladder operator method of
Dirac, which was later extended to all other solvable problem under the rubric of
supersymmetric quantum mechanics (SUSY)\cite{susyqm}.

\sms In supersymmetric quantum mechanics, the partner potentials $V_{\pm}(x)$ are
related to the superpotential $W(x)$ by $V_{\pm}(x)=W^2(x) \pm W'(x)$. Setting
$\hbar=m=1$, the corresponding Hamiltonians $H_{\pm}$ have a factorized form
\be \label{hpm}
H_-=\Ap \Am~,~H_+=\Am \Ap~,~\Am \equiv
{1\over \sqrt{2}}\({d \over {dx}} +W\)~,~
\Ap \equiv {1\over \sqrt{2}}\(-{d \over {dx}}
+W\)
\ee
If either ${\psi_0^{(-)}}(x) \equiv \exp\left( - \int^x W(y) dy \right)$ or $
\frac{1}{\( {\psi_0^{(-)}}(x)\)^{-1}}$ is normalizable, it is the ground state for
$H_-$ or $H_+$ and corresponds to the case of unbroken SUSY. As can be explicitly
checked, $\Am \psi_0^{(-)} = 0$, and thus the ground state $E_0^{(-)}$ of $H_-$ is
zero. The vanishing of the ground state energy of Hamiltonian $H_-$ or $H_+$, is a
necessary and sufficient condition for unbroken SUSY \cite{susyqm}.
If, however, neither
${\psi_0^{(-)}}$ nor $\frac{1}{{\psi_0^{(-)}}}$   is well defined, the SUSY is said to
be spontaneously broken. In that situation, the Hamiltonians $H_+$ and $H_-$ have
exactly the same eigenvalues and neither has a zero energy groundstate.

 Assuming, as in Ref.
\cite{das} that $H_+$ has the zero energy groundstate, the eigenstates of $H_+$ and
$H_-$ are related by
\be \label{SUSY}
E^{(+)}_0 = 0~,~~~
E^{(+)}_{n+1}\;=\; E^{(-)}_{n},~~~
\psi^{(-)}_{n}\propto \Ap \,\psi^{(+)}_{n+1}~~,~~~
\Am \,\psi^{(-)}_{n} \propto \psi^{(+)}_{n+1}  ~~, \quad   n=0,1,2, \ldots. \ee

\sss
\noindent {\large{\bf Supersymmetry in singular systems}}

\ni We have studied oscillator systems with a $\frac{1}{x^2}$ singularity
\cite{eqshift,singular_pot,brknsusy}. It was explicitly shown that, despite the
presence of the singularity, supersymmetry remains unbroken, provided the
superpotential is regularized with care, in agreement with the conclusions of Ref.
\cite{das}. In Ref. \cite{brknsusy}, we showed how to determine the spectrum
algebraically if supersymmetry is broken. Here, we would like to revisit the
half-oscillator system and check whether it can support an unbroken supersymmetry.
Following the notation used in Ref.\cite{das}, we define the half-oscillator by the
potential:
\be
 V(x)= \left\{
    \begin{array}{cc}
         \infty  ~~~~~~~& x<0\\
        {1 \over 2} ~\(\omega^2 x^2 -\omega\)
        ~~~~~~~& x>0
    \end{array}
\right. \ee

\ni If the potential ${1 \over 2} \(\omega^2 x^2 -\omega\)$ defined over the positive
part of the $x$-axis were to be extended to the entire real axis, the system would have
a zero energy groundstate and unbroken SUSY. This potential is generated by the
following superpotential
\be \label{superpotential}
 W(x)= \left\{
    \begin{array}{cc}
         \infty  ~~~~~~~& x<0\\
        - ~\omega x
        ~~~~~~~& x>0
    \end{array}
\right.
\ee
 As shown in Ref. \cite{das,eqshift,singular_pot,brknsusy}, the ground state
 energy of this model is nonzero, signalling the
 breakdown of the supersymmetry. It was argued in Ref.\cite{das} that this breaking of SUSY is
 actually an artifact of the way one defines the potential for the negative value of $x$, and thus,
 if the superpotential is chosen to
be $c$, a positive constant, in the region $x<0$ and $- ~\omega x ~{\rm for
}~x>0$, and a careful limit $c\rightarrow \infty$ is taken, that would lead to a zero
energy groundstate for
$H_+$. I.e., the claim is that it is not the singularity itself, but rather the way it
is handled, that results in the violation of the supersymmetry. We would like to check
this assertion.

\sss
\ni
We start with the superpotential of Ref. \cite{das}:
%
\be W(x,c)= c~\theta(-x)- \omega\, x\, \theta(x)\ee
where $\theta(x)$ is the Heavyside step function. Corresponding partner potentials
$V_\pm$ are given by
\be
V_\pm ={1 \over 2} \[ c^2~\theta(-x) +
\( \omega^2 x^2  \mp  \omega \) \theta(x)
\mp c\delta(x)
\]
\ee \ni where $D\(\frac{\epsilon}{\omega}, \sqrt{2\omega\,x}\)$ is the parabolic
cylindrical function \cite{Merzbacher}. The finiteness of the potential on the negative $x$-axis extends the domain to
($-\infty,\infty$). For both potentials, solutions away from the origin are given by:
\br
x>0 & ~~~~~~\psi(x) = &A_1 D\(\frac{\epsilon}{\omega}~, \sqrt{2\omega}\,x \) \nn\\
x<0 & ~~~~~~\psi(x) = &A_2  \exp\(\sqrt{C^2-2\epsilon}~x\) \er \ni where
$D\(\frac{\epsilon}{\omega}, \sqrt{2\omega\,x}\)$ is the parabolic cylindrical function
\cite{Merzbacher}. The bound states are determined from the boundary conditions:
\be \label{bc-} \psi(0_-) =\psi(0_+) ~~{\rm and}~~ \psi'(0_+) -
\psi'(0_-)=-c\psi(0_-)\ee
\sss \ni where $D\(\frac{\epsilon}{\omega}, \sqrt{2\omega\,x}\)$ is the parabolic
cylindrical function \cite{Merzbacher}. Using
\be D\(\frac{\epsilon}{\omega}~,~ 0\) = \frac{2^{\frac{\epsilon}{2\omega}}\sqrt{\pi}}
{~\Gamma\({1\over 2}-\frac{\epsilon^+}{2\omega}\)} ~~{\rm and }~~
D'\(\frac{\epsilon}{\omega}~,~ 0\) =
 -2 \frac{2^{\frac{\epsilon}{2\omega}}\sqrt{\pi}~\sqrt{\omega}}
 {\Gamma\(-\frac{\epsilon}{2\omega}\)}
~~~~~~~~~~~~~~~~~~~~~~~~~~~~~ \ee
and the boundary conditions given in eq. $\!$(\ref{bc-}), we get\\
\be \label{eqa}
-2
\frac{\sqrt{\omega}~~\Gamma\(\frac{1}{2}-\frac{\epsilon^+}{2\omega}\)}
{\Gamma\(-\frac{\epsilon^+}{2\omega}\)}
=
{\sqrt{c^2-2\epsilon^+}-c}\ee
which agrees with eq. $\!\!$(23) of Ref. \cite{das}. For any finite $c$, it is easy to
show that  $\epsilon^+$ = 0 is a solution. Using $\Gamma(x+1)=x~\Gamma(x)$, the left
hand side may be rewritten as
$-\frac{{\epsilon^+}~\Gamma\(\frac{1}{2}-\frac{\epsilon^+}{2\omega}\)}
{\sqrt{\omega}~~\Gamma\(1-\frac{\epsilon^+}{2\omega}\)}$. Since both $\Gamma$ functions
are finite as $\epsilon^+ \rightarrow 0$, eq. $\!$(\ref{eqa}) is manifestly satisfied
by the solution $\epsilon^+=0$. Thus, for any finite value of $c$, a zero energy
eigenvalue is indeed a solution of the above equation; therefore, the supersymmetry
remains unbroken. Other eigenvalues $0< \epsilon^+\!<c^2$, for a finite $c$, can be
obtained by numerically solving eq. $\!$(\ref{eqa}).

\sss \ni Now let us consider the case of half-oscillator; i.e., take the limit $c
\rightarrow \infty$.  For $\epsilon^+<<c^2$, we have, inverting eq. $\!$(\ref{eqa}),
\br  \label{eqb} \frac
{\sqrt{\omega}~~\Gamma\(1-\frac{\epsilon^+}{2\omega}\)}
{\epsilon^+~\Gamma\({1\over 2}-\frac{\epsilon^+}{2\omega}\)} \eq - {c\over
{\epsilon^+}}\er
For $\epsilon^+\approx 0$, the above equation reduces to \br \frac
{\sqrt{\omega}}
{\epsilon^+~\sqrt{\pi}} \eq - {c\over {\epsilon^+}}\nn\er
\ni Both sides diverge as $1\over \epsilon^+$; therefore, for finite $\omega$, this
does not permit infinite  values of $c$. Thus, in contrast to the results obtained in
Ref. \cite{das}, we find that  $\epsilon^+=0$ is not a permissible solution. To find
the correct eigenenergies of this system as $c\rightarrow \infty$, we write eq.
$\!$(\ref{eqb}) as
\begin{equation}
\frac {\sqrt{\omega}~~\Gamma\(1-\frac{\epsilon^+}{2\omega}\)}
{\Gamma\({1\over 2} -\frac{\epsilon^+}{2\omega}\) } =- {c} \ee

\ni which yields $\(1-\frac{\epsilon^+}{2\omega}\)=-n_+$ with $n_+ =0, 1, 2,  \cdots$;
i.e., $\epsilon^+_{n_+}= 2 \omega(n_+ + 1)$. In summation, the ground state energy for
$V_+$ is not equal to 0 but rather to $2\omega$.

\sss \ni For the partner potential $V_-$, the eigenenergies are determined by
imposition of boundary conditions: \be \label{bc} \psi(0_-) =\psi(0_+) ~~{\rm and}~~
\psi'(0_+) - \psi'(0_-)=c\psi(0_-)\ee

\ni which gives rise to the condition
 \be \label{eq9} -\frac{1}{2}
 \frac{\Gamma\(\frac{1}{2}-\frac{\epsilon^-}
 {2\omega}\)}{\sqrt{\omega}~~\Gamma\( 1-\frac{\epsilon^-}{2\omega}\)}
 = %
 \frac{1}{\sqrt{c^2-2\epsilon^-}+c}\ee
\ni and by similar arguments, the eigenenergies are given by $\epsilon^-_{n_-}= 2
\omega(n_- + 1)$, $n_- = 0, 1, 2,  \cdots$. Thus, when we take the $c\rightarrow
\infty$ limit, the spectrum of systems with potentials $V_+$ and $V_-$ are identical,
characteristic of broken supersymmetry.

\sss
\ni
An alternate regularization considered in Ref. \cite{das} is
\br &W(x) =  & -\omega\, x ~\theta(x) - \lambda\, x ~\theta(-x) \nn\\
&V_\pm ~~~~=&{1 \over 2} \( \omega^2 x^2  \mp  \omega \) ~\theta(x) + {1 \over 2}\(
\lambda^2 x^2 \mp  \lambda \) ~\theta(-x) \label{eqz} \er
Boundary conditions lead to
\be \label{eqx}
\frac
{\epsilon^+~\Gamma\({1\over 2}-\frac{\epsilon^+}{2\omega}\)}
{\sqrt{\omega}~~\Gamma\(1-\frac{\epsilon^+}{2\omega}\)} =
\frac{\epsilon^+~\Gamma\({1\over 2}-\frac{\epsilon^+}{2\lambda}\)}
{\sqrt{\lambda}~~\Gamma\(1-\frac{\epsilon^+}{2\lambda}\)}
 \ee \ni  For finite $\lambda$, the above equation obviously
 admits a zero energy solution and hence has unbroken supersymmetry.
 Other eigenvalues can be obtained by solving the above equation
 numerically for a fixed $\lambda$.

 \ss
 \ni
 For $\lambda\rightarrow
\infty$, however, analysis identical to the preceding yields
\be \label{eqy}
\frac{\sqrt{\omega}}{\epsilon^+}\frac{\Gamma\(1-\frac{\epsilon^+}{2\omega}\)}{~~\Gamma\({1\over
2}-\frac{\epsilon^+}{2\omega}\)} =
\frac{\sqrt{\lambda}}{\epsilon^+}\frac{\Gamma\(1-\frac{\epsilon^+}{2\lambda}\)}{~~\Gamma\({1\over
2}-\frac{\epsilon^+}{2\lambda}\)} \ee
In the limit $\epsilon^+\rightarrow 0$, the left hand side goes to
$\frac{\sqrt{\omega}}{\epsilon^+\sqrt{\pi}}$ while the right hand side goes to
$\frac{\sqrt{\lambda}}{\epsilon^+\sqrt{\pi}}$.  Thus, for finite $\omega$ this does not
allow for $\lambda\rightarrow\infty$. Therefore, for this regularization as well, as
$\lambda\rightarrow \infty$, $\Gamma\(1-\frac{\epsilon^+}{2\omega}\)$ needs to be in
the neighborhood of its poles, i.e., $\(1-\frac{\epsilon^+}{2\omega}\)=-n_+$, i.e.,
$\epsilon^+_{n_+}= 2 \omega(n_+ + 1)$. With similar arguments for $\epsilon^-$,  we get
$\epsilon^\pm_n= 2 \omega(n + 1)$, where $n=0,1,2, \cdots$. Identicality of the two
spectra means that supersymmetry is broken.

\sss \ni {\bf Conclusion:}~We revisited the analysis of supersymmetry in an oscillator
in presence of a singularity. While we agree with the assertion of Ref. \cite{das} that
systems with $1\over x^2$, if properly regularized, retain supersymmetry
\cite{eqshift,singular_pot,brknsusy}, the half-oscillator system does not. We have
explicitly shown that a careful analysis leads to identical spectra for both partner
potentials of the half-oscillator system, and hence to broken supersymmetry.

\sss \ni {\bf Acknowledgment:}~We would like to thank Prof. R. Dutt and Prof.
Constantin Rasinariu for carefully reading the manuscript and suggesting changes. We
also thank Prof. A. Khare and Prof. U.P. Sukhatme for important discussion. One of us
(AG) would like to thank the physics department at UIC for its warm hospitality.


\end{document}